\documentclass[12pt]{article}
\usepackage{amsmath,epsfig}

\numberwithin{equation}{section}
\renewcommand{\thefootnote}{\fnsymbol{footnote}}

\begin{document}

\title{
\begin{flushright}
\ \\*[-80pt] 
\begin{minipage}{0.25\linewidth}
\normalsize
hep-th/0702097 \\
YITP-07-06 \\
OU-HET 574/2007 
\\*[50pt]
\end{minipage}
\end{flushright}
{\Large \bf 
Supersymmetry breaking in a warped slice 
with Majorana-type masses
\\*[20pt]}}

\author{Hiroyuki~Abe$^{1,}$\footnote{
E-mail address: abe@yukawa.kyoto-u.ac.jp} \ and \ 
Yutaka~Sakamura$^{2,}$\footnote{
E-mail address: sakamura@het.phys.sci.osaka-u.ac.jp} \\*[20pt]
$^1${\it \normalsize 
Yukawa Institute for Theoretical Physics, Kyoto University, 
Kyoto 606-8502, Japan} \\
$^2${\it \normalsize 
Department of Physics, Osaka University, 
Toyonaka, Osaka 560-0043, Japan} \\*[50pt]}

\date{
\centerline{\small \bf Abstract}
\begin{minipage}{0.9\linewidth}
\medskip 
\medskip 
\small
We study the five-dimensional (5D) supergravity compactified on 
an orbifold $S^1/Z_2$, where the $U(1)_R$ symmetry is gauged by 
the graviphoton with $Z_2$-even coupling. 
In contrast to the case of gauging with $Z_2$-odd coupling, 
this class of models has Majorana-type masses and allows 
the Scherk-Schwarz (SS) twist even in the warped spacetime. 
Starting from the off-shell formulation, we show that the 
supersymmetry is always broken in an orbifold slice of AdS$_5$, 
{\it irrespective of} the value of the SS twist parameter. 
We analyze the spectra of gaugino and gravitino in such 
background, and find the SS twist can provide sizable 
effects on them in the small warping region. 
\end{minipage}
}

\begin{titlepage}
\maketitle
\thispagestyle{empty}
\end{titlepage}

\clearpage

\renewcommand{\thefootnote}{\arabic{footnote}}
\setcounter{footnote}{0}

\section{Introduction}
The five-dimensional (5D) warped spacetime has various aspects 
for the physics beyond the standard model from the view point of 
both the phenomenological and the theoretical studies. For example, 
a hierarchy between the weak and the Planck scale and ones among the 
observed quark/lepton masses and mixings can be explained by the 
localization of graviton~\cite{Randall:1999ee} and matter fields~\cite{
Arkani-Hamed:1999dc} respectively in an orbifold slice of the 5D 
anti-de Sitter space, AdS$_5$. 
In the theoretical side, four-dimensional (4D) strongly coupled 
gauge theories can be studied within weakly coupled theories 
in AdS$_5$ by the use of so-called AdS/CFT 
correspondence~\cite{Maldacena:1997re}. 
Moreover, it is known that a low energy effective theory of strongly 
coupled heterotic string~\cite{Horava:1995qa} can be described by 5D 
supergravity (SUGRA) on a curved background~\cite{Lukas:1998yy}. 

Among these features, the structure of supersymmetry 
(SUSY)~\cite{Altendorfer:2000rr,Gherghetta:2000qt,Luty:2000ec} 
in an orbifold slice of AdS$_5$, and the mediation patterns of 
SUSY breaking from the hidden to the visible sector~\cite{Marti:2001iw} 
in such background is particularly interesting. Here we focus on some 
phenomenological models in a slice of AdS$_5$ with local SUSY. 
In the SUGRA framework, the $U(1)_R$ gauging is necessary to 
generate a bulk cosmological constant which makes the background warped. 
Then we have basically two ways for realization of such warped geometry, 
which are distinguished from each other by the orbifold $Z_2$-parity 
of the gauge coupling constant for the $U(1)_R$ symmetry gauged by the 
$Z_2$-odd graviphoton. 

In the case of $Z_2$-odd $U(1)_R$ gauge coupling, an on-shell 
SUGRA-matter-Yang-Mills system was constructed~\cite{Gherghetta:2000qt}, 
and a lot of phenomenological studies have been done so far 
(see, e.g., \cite{Gherghetta:2000kr}). However, in order to embed 
such nontrivial $Z_2$-odd coupling constant into SUGRA, which changes 
the sign across the orbifold boundaries, we need a four-form 
Lagrange multiplier~\cite{Bergshoeff:2000zn}. 
On the other hand, in the $Z_2$-even coupling case, almost only the 
on-shell formulation for pure SUGRA case has been 
studied~\cite{Altendorfer:2000rr}, even though it seems simpler 
than the $Z_2$-odd coupling case in the sense that it does not 
require any Lagrange multiplier whose origin is unclear within 
the framework of 5D SUGRA.

In this paper, we study the latter case, i.e., 5D SUGRA compactified 
on an orbifold $S^1/Z_2$ where the $U(1)_R$ symmetry is gauged by 
the graviphoton with {\it $Z_2$-even} coupling, starting from the off-shell 
formulation~\cite{Fujita:2001bd,Kugo:2002js}. The off-shell formulation 
allows us to construct such SUGRA model coupled to an arbitrary number of 
matter and gauge fields even with boundary terms. 
Interestingly enough, this model allows the Scherk-Schwarz (SS) 
twist~\cite{Scherk:1978ta} even in the warped geometry~\cite{Abe:2005wn}, 
in contrast to the case of $Z_2$-odd $U(1)_R$ coupling where it is 
prohibited from the consistency of the theory~\cite{Hall:2003yc}. 
In the absence of the SS twist, it is briefly shown in Ref.~\cite{Kugo:2002js} 
that SUSY is spontaneously broken in a slice of AdS$_5$ in the case of 
$Z_2$-even coupling. We reexamine this fact in more general setup, i.e., 
including the SS twist, and obtain a result that SUSY is always broken 
{\it irrespective of the value of the SS twist parameter}. 

A scenario of SS SUSY breaking has been extensively studied in a 
flat space. In such a case, the SS twist can be interpreted as a 
Wilson line for some auxiliary gauge field in the off-shell 
formulation~\cite{Kaplan:2001cg,vonGersdorff:2001ak}, and appears 
as a radion mediated SUSY breaking~\cite{Chacko:2000fn} 
in the 4D effective theory. Thus it is not an explicit breaking of SUSY and 
gives a controllable result. Furthermore we have an interesting correlation 
between the flavor structures of fermion and sfermion when the hierarchies 
among quark/lepton masses are consequence of the localization of matter 
fields~\cite{Abe:2004tq}. 
However the effect of the SS twist on the soft SUSY breaking parameters 
in the warped geometry has not been investigated yet. So we study the 
structure of SUSY breaking in detail. 

The sections of this paper are organized as follows. 
In Section~\ref{sec:z2even}, we show the Lagrangian for the 5D gauged 
SUGRA with $Z_2$-even coupling, based on the off-shell formulation. 
For generality, we also include the SS twist parameter in the action. 
In Section~\ref{sec:killing}, we analyze the Killing spinor equations 
and show that SUSY is spontaneously broken in a slice of AdS$_5$ 
{\it irrespective of} the value of the twist parameter. 
In Section~\ref{sec:masses}, we study SUSY breaking structure 
in detail by computing spectra of the gaugino and the gravitino. 
Section~\ref{sec:conclusion} is devoted to summary and discussions. 
In Appendix~\ref{app:offshell}, we briefly review the off-shell 
formulation of 5D SUGRA on orbifold derived in 
Refs.~\cite{Fujita:2001bd,Kugo:2002js}. 

\section{5D gauged supergravity with $Z_2$-even coupling}
\label{sec:z2even}
In this section, we show the Lagrangian for 5D SUGRA compactified 
on an orbifold $S^1/Z_2$ where the $U(1)_R$ symmetry is gauged by 
the graviphoton with a $Z_2$-even gauge coupling. We utilize the 
off-shell formulation given in Refs.~\cite{Fujita:2001bd,Kugo:2002js} 
and reviewed in Appendix~\ref{app:offshell}, in order to include 
physical $n_V$ vector multiplets and $n_H$ hypermultiplets with 
some superpotential terms at the orbifold boundaries. 

The ingredients of the off-shell formulation for our setup 
is the Weyl multiplet, ($e_\mu^{\ m}$, $\psi_\mu^i$, $V_\mu^{ij}$, 
$b_\mu$, $v^{mn}$, $\chi^i$, $D$), the vector multiplets, 
($M$, $W_\mu$, $\Omega^i$, $Y^{ij}$)$^I$ and the hypermultiplets, 
(${\cal A}^\alpha_{\ i}$, $\zeta^\alpha$, ${\cal F}^\alpha_{\ i}$). 
The index $i=1,2$ is the 
$SU(2)_{\mbox{\scriptsize \boldmath $U$}}$-doublet index.
The Greek subscripts $\mu$, $\nu$, $\ldots$ describe the vector 
indices for 5D curved space, and the corresponding tangent flat 
space indices are represented by Roman subscripts $m$, $n$, $\ldots$. 
These subscripts with under-bar will be used for the 4D space 
other than the compact fifth dimension $\mu=y$ or $m=4$. 

The index $I= 0,1,2,\ldots,n_V$ labels the vector multiplets, 
and the $I=0$ multiplet is introduced to yield graviphoton degree of 
freedom which is included in the on-shell SUGRA (Weyl) multiplet. 
For hyperscalars ${\cal A}^\alpha_{\ i}$, the index $\alpha$ runs over 
$\alpha=1,2,\ldots,2(n_C+n_H)$ where $n_C$ and $n_H$ are the numbers 
of the compensator and the physical hypermultiplets, respectively. 
Each hyperscalar has a quaternionic structure, and is written as 
\begin{eqnarray}
\left( 
\begin{array}{cc}
{\cal A}^{2\hat\alpha-1}_{\ i=1} 
& {\cal A}^{2\hat\alpha-1}_{\ i=2} \\
{\cal A}^{2\hat\alpha}_{\ i=1} 
& {\cal A}^{2\hat\alpha}_{\ i=2} 
\end{array}
\right) 
&=& 
\left( 
\begin{array}{cc}
({\cal A}^{\hat\alpha}_+)^{\ast} 
& {\cal A}^{\hat\alpha}_- \\
({\cal A}^{\hat\alpha}_-)^{\ast} 
& {\cal A}^{\hat\alpha}_+ 
\end{array}
\right), 
\label{eq:quaternion}
\end{eqnarray}
where $\hat\alpha=1,\ldots,n_C+n_H$. 
In this paper we adopt the single compensator case, 
$n_C=1$, and divide the indices into two pieces such as 
$\alpha=(a,\underline\alpha)$ where $a=1,2$ and 
$\underline\alpha=1,2,\ldots,2n_H$ 
are indices for the compensator and the physical hypermultiplets, 
respectively. The $Z_2$-parity assignments for the fields in the 
above multiplets are summarized in Table~\ref{tab:parity} in 
Appendix~\ref{app:offshell}. Throughout this paper, we work in a 
unit of the 5D Planck mass, $M_5=1$. 

The most general off-shell action for 5D SUGRA on $S^1/Z_2$ 
orbifold is given by~\cite{Fujita:2001bd} 
\begin{eqnarray}
S &=& \int d^4x \int dy\, 
\bigg\{ {\cal L}_b + {\cal L}_f + {\cal L}_{\rm aux} 
+\sum_{l=0,\pi} {\cal L}^{(l)} \delta(y-lR) \bigg\}, 
\end{eqnarray}
where ${\cal L}_b$, ${\cal L}_f$ and ${\cal L}_{\rm aux}$ are the 
Lagrangians for the bosonic, fermionic and auxiliary fields, 
respectively. The ${\cal L}^{(l)}$ stands for the boundary 
Lagrangian. These are explicitly shown in Eqs.~(\ref{eq:bulklag}) 
and (\ref{eq:bdrylag}). 

\subsection{$U(1)_R$ gauging and tensions at boundaries}
\label{ssec:gauging}
One of important points to realize the warped background geometry 
is that cosmological constant terms are included in the Lagrangian, 
otherwise the background becomes flat. In order to obtain an orbifold 
slice of AdS$_5$, 
\begin{eqnarray}
ds^2 &=& 
e^{-2 \sigma(y)} \eta_{\underline\mu \underline\nu} 
dx^{\underline\mu} dx^{\underline\nu} -dy^2, 
\quad \sigma(y) \ = \ k|y|, 
\label{eq:ads5}
\end{eqnarray}
as a solution of the Einstein equation, 
the bulk cosmological constant $\Lambda$ must be balanced 
with the tensions $\Lambda^{(0)}$, $\Lambda^{(\pi)}$ 
at the boundaries~\cite{Randall:1999ee}. 
This is the so-called Randall-Sundrum (RS) relation, 
\begin{eqnarray}
\sqrt{-6\Lambda} &=& \Lambda^{(0)} \ = \ -\Lambda^{(\pi)}. 
\label{eq:rs}
\end{eqnarray}
The bulk cosmological constant is generated in SUGRA framework by 
gauging $R$-symmetry with a $Z_2$-odd gauge field. 
The most economical choice for such gauge 
field is the graviphoton which always exists in 5D SUGRA. 

In the off-shell (superconformal) formulation of 5D SUGRA, 
the $SU(2)_R$ symmetry is realized as a diagonal subgroup of 
$SU(2)_{\mbox{\scriptsize \boldmath $U$}} \times SU(2)_C$, 
where $SU(2)_{\mbox{\scriptsize \boldmath $U$}}$ is a gauge 
symmetry of superconformal group\footnote{The corresponding 
gauge field ${V_\mu}^i_{\ j}$ is an auxiliary field and 
thus non-dynamical.} which rotates the index $i=1,2$, 
and $SU(2)_C$ is an isometry group which rotates only the 
compensator index $a=1,2$ (see the $\mbox{\boldmath $U$}$ 
gauge-fixing condition in Eq.~(\ref{eq:scgf})). 
The simplest version of such $R$-gauging is a gauging of 
$U(1)$ subgroup of $SU(2)_R$. 
In this case, the covariant derivative of 
the compensator field is given by 
\begin{eqnarray}
\nabla_\mu {\cal A}^a_{\ i} &=& 
\partial_\mu {\cal A}^a_{\ i} 
-B_\mu g_R (t_R)^a_{\ b} {\cal A}^b_{\ i} 
+\cdots, 
\end{eqnarray}
with 
\begin{eqnarray}
t_R &=& i \vec{q} \cdot \vec\sigma, 
\qquad |\vec{q}|=1, 
\label{eq:gauging}
\end{eqnarray}
where $B_\mu$ is the graviphoton, 
$\vec{q}$ is a three-dimensional unit vector 
that indicates the direction of the gauging, and 
$\vec\sigma=(\sigma_1,\,\sigma_2,\,\sigma_3)$ 
are Pauli matrices. 
In general, the orbifold projection is expressed as 
\begin{eqnarray}
\Phi^i(x,-y) &=& \Pi Z^i_{\ j} \Phi^j(x,y). 
\end{eqnarray}
where $\Pi=\pm 1$ is the $Z_2$-parity eigenvalue of 
the field $\Phi$. Without loss of generality, 
we choose $Z=\sigma_3$ in the following. 
The $Z_2$-parity of compensator is assigned such that the 
diagonal components are even while the off-diagonal ones are odd 
in Eq~(\ref{eq:quaternion}). 
(See Table~\ref{tab:parity} in Appendix~\ref{app:offshell}.) 

Since the graviphoton is $Z_2$-odd, the gauge coupling $g_R$ 
associated to the gauging (\ref{eq:gauging}) has to be $Z_2$-odd 
for $\vec{q}=(0,\,0,\,1)$ while $Z_2$-even for 
$\vec{q}=(q_1,\,q_2,\,0)$. Most of the phenomenological studies 
in a slice of AdS$_5$ have been done in the case of 
$\vec{q}=(0,\,0,\,1)$ with $Z_2$-odd coupling 
$g_R \propto \epsilon(y)$ where $\epsilon(y)$ is the periodic 
sign function of the compactified extra coordinate $y$. 
Such $Z_2$-odd coupling can be obtained in a sense dynamically 
by introducing four-form Lagrange multiplier~\cite{Bergshoeff:2000zn} 
into the off-shell formulation~\cite{Fujita:2001bd}. 
This multiplier simultaneously generates tensions at the 
boundaries proportional to 
$\partial_y \epsilon(y)=2(\delta(y)-\delta(y-\pi R))$, 
which automatically satisfies the RS relation~(\ref{eq:rs}). 
Here $R$ is the radius of the orbifold. 
Thus, this SUGRA model admits an orbifold slice of AdS$_5$ as 
a background solution of the Einstein equation~\cite{Fujita:2001bd}. 
In fact, this background preserves N=1 SUSY, and thus it realizes 
a SUSY RS model whose on-shell formulation is provided in 
Ref.~\cite{Gherghetta:2000qt}. 

On the other hand, we do not need such Lagrange multiplier 
in 5D SUGRA framework in the case of $\vec{q}=(q_1,\,q_2,\,0)$ with 
$Z_2$-even $g_R$. However, in this case, we have to find some mechanism 
to generate the tensions at the boundaries because the contribution 
to the tensions from the Lagrange multiplier no longer exists. In the 
off-shell formulation, constant pieces in the superpotential at the 
orbifold boundaries generate such tensions because now the $U(1)_R$ 
symmetry is gauged with $Z_2$-even coupling $g_R$ which does not vanish 
at the boundaries\footnote{
This contribution to the tensions does not exist 
in the case of gauging with $Z_2$-odd coupling 
because $g_R$ vanishes at the boundaries.}. 

Thus we introduce $N=1$ invariant constant superpotential terms at 
the orbifold boundaries, 
\begin{eqnarray}
\sum_{l=0,\pi}{\cal L}^{(l)} \delta(y-lR) 
&\supset& {\cal L}_W \ = \ 
\sum_{l=0,\pi} \delta(y-lR) [\phi^3 W^{(l)}]_F, 
\label{eq:csp}
\end{eqnarray}
where $\phi$ is a compensator chiral multiplet 
induced from the bulk compensator hypermultiplet, 
whose lowest component is ${\cal A}^{a=2}_{\ i=2}$ and 
$[\ldots]_F$ represents the $F$-term invariant formula 
of $N=1$ superconformal tensor calculus~\cite{Kugo:2002js,Kugo:1982cu}. 
The complex constant $W^{(l)}$ is parameterized as 
$W^{(l)}=w^{(l)}_2+iw^{(l)}_1$ 
with real constants $w^{(l)}_{1,2}$. 
Because ${\cal L}_W$ contains a tadpole 
of the auxiliary field ${\cal F}^a_{\ i}$ 
in the compensator hypermultiplet (as well as of $V_y^{ij}$), 
we rearrange the ${\cal F}$-related part of auxiliary 
Lagrangian ${\cal L}_{\rm aux}$ shown in Eq.~(\ref{eq:bulklag}) 
and find 
\begin{eqnarray}
{\cal L}_W + {\cal L}_{\rm aux}^{{\cal F}} 
&=& {\cal L}_{\rm tension} 
+{\cal L}_w +{\cal L}_{\rm aux}^{{\cal F}} 
\big|_{{\cal F}^a_{\ i} \to {\cal F}^a_{\ i}+C^a_{\ i}}, 
\end{eqnarray}
where 
\begin{eqnarray}
{\cal L}_{\rm tension} &=& 
-4e_{(4)}g_R M^{I=0} \sum_{l=0,\pi} \delta(y-lR) 
\big( q_1 w^{(l)}_2 - q_2 w^{(l)}_1 \big) 
\big| {\cal A}^{a=2}_{\ i=2} \big|^2, 
\\
{\cal L}_w &=& 
e_{(4)} \sum_{l=0,\pi} \delta(y-lR) (w^{(l)}_2+iw^{(l)}_1) 
\bigg[ 2{\cal A}^{a=2}_{\ i=2} 
\big( \nabla_4{\cal A}+{\cal A}V_4 
-2i\bar\psi_4 \zeta \big)^{a=2}_{\ i=1} 
\nonumber \\ &&
-2\bar\zeta_+^{\hat\alpha=1} 
{\cal P}_R \zeta_+^{\hat\alpha=1} 
+4i{\cal A}^{a=2}_{\ i=2} 
\bar\psi_+ \cdot \gamma {\cal P}_R \zeta_+^{\hat\alpha=1} 
-2({\cal A}^{a=2}_{\ i=2})^2 
\bar\psi_{m+} \gamma^{mn} {\cal P}_L \psi_{n+} \bigg] 
+ {\rm h.c.} 
\nonumber \\ &&
+2e \bigg( e^{-1}e_{(4)} \sum_{l=0,\pi} \delta(y-lR) 
\big| \vec{w}^{(l)} \big| \bigg)^2 
({\cal A}^{a=2}_{\ i=2})^2, 
\\
C^a_{\ i} &=& 
e^{-1}e_{(4)} \sum_{l=0,\pi} \delta(y-lR) 
\left( \frac{(M^{I=0})^{-1} W_4^{I=0} \mathbf{1}_2 + {\sigma_3}}
{1+(M^{I=0})^{-2}(W_4^{I=0})^2} \right)^a_{\ b} 
(i\vec{w}^{(l)} \cdot \vec\sigma)^b_{\ c} {\cal A}^c_{\ i}. 
\end{eqnarray}
Here $e_{(4)}$ is a determinant of the induced 4D vielbein 
and $\vec{w}^{(l)}=(w_1^{(l)},w_2^{(l)},0)$. 
Then, the total Lagrangian is obtained as 
\begin{eqnarray}
{\cal L} 
&=& {\cal L}_b + {\cal L}_f 
+{\cal L}_{\rm tension} 
+{\cal L}_w +{\cal L}_{\rm aux} 
\big|_{{\cal F}^a_{\ i} \to {\cal F}^a_{\ i}+C^a_{\ i}}. 
\label{eq:untwistedlagrangian}
\end{eqnarray}
Therefore the boundary tension is proportional to the constant 
$w_{1,2}^{(l)}$ and also the gauge coupling $g_R$. 
We successfully obtain an orbifold slice of AdS$_5$ 
background if we tune the complex constants $W^{(l)}$ to 
suitable values satisfying the RS relation (\ref{eq:rs}). 

As pointed out in Ref.~\cite{Kugo:2002js}, any value of $W^{(l)}$ 
does not allow a Killing spinor in 4D Poincar\'e invariant vacuum. 
This means that SUSY is spontaneously broken in the slice of 
AdS$_5$, if we choose the off-diagonal $U(1)_R$-gauging 
$\vec{q}=(q_1,q_2,0)$ with $Z_2$-even $g_R$. 
This fact was originally pointed out in Ref.~\cite{Zucker:2000ks} 
in a framework of linear multiplet compensator formalism. 
Here, notice that this class of theory admits the Scherk-Schwarz (SS) 
twist even in a warped spacetime~\cite{Abe:2005wn}. Thus, there might 
be a possibility that the twist recovers SUSY in a slice of AdS$_5$. 
We examine this possibility in Section~\ref{sec:killing}, 
and show that any value of twist parameter cannot recover SUSY. 

\subsection{Scherk-Schwarz twist}
\label{ssec:twist}
Before analyzing the Killing spinor equations, we show how to 
include the SS twist into the previous SUGRA model 
described by the Lagrangian~(\ref{eq:untwistedlagrangian}). 

When we consider the torus compactification of fifth dimension, 
we have a nontrivial physical parameter (even in the pure SUGRA) 
called the SS twist~\cite{Scherk:1978ta} or the Wilson 
line associated to the $SU(2)_{\mbox{\scriptsize \boldmath $U$}}$ 
symmetry~\cite{vonGersdorff:2001ak}. 
Such a physical parameter can be introduced in the framework of 
5D conformal SUGRA in the following way~\cite{Abe:2005wn}. 

The normal (untwisted) $SU(2)_{\mbox{\scriptsize \boldmath $U$}}$ 
gauge fixing for the compensator scalar is given by 
\begin{eqnarray}
{\cal A}^a_{\ i} 
\equiv \delta^a_{\ i} 
\sqrt{1+{\cal A}^{\underline\alpha}_{\ j} 
{\cal A}_{\underline\alpha}^{\ j}}. 
\label{eq:utuf}
\end{eqnarray}
The SS twist can be incorporated into the theory by 
modifying this condition as 
\begin{eqnarray}
{\cal A}^a_{\ i} \equiv 
\left( e^{i\vec\omega \cdot \vec\sigma\,f(y)} \right)^a_{\ i} 
\sqrt{1+{\cal A}^{\underline\alpha}_{\ j} 
{\cal A}_{\underline\alpha}^{\ j}}, 
\label{eq:tuf}
\end{eqnarray}
where $\vec\omega=(\omega_1,\omega_2,\omega_3)$ 
is the twist vector which determines the magnitude 
and the direction of the twisting, and $f(y)$ is 
a gauge fixing function satisfying 
$f(y+\pi R)=f(y)+\pi$.\footnote{
The function $f(y)$ can be chosen arbitrarily 
as long as it satisfies this condition. 
The simplest choice is $f(y)=y/R$.}
As shown in Ref.~\cite{Abe:2005wn}, the twist is possible 
only when the twisting direction is the same as the 
$U(1)_R$ gauging direction, 
$\left[ (\vec{q} \cdot i\vec\sigma),\, 
(\vec{\omega} \cdot i\vec\sigma) \right]=0$, i.e., 
\begin{eqnarray}
\vec\omega &=& \omega\,(q_1,q_2,0), 
\label{eq:ccfrg}
\end{eqnarray}
otherwise the resulting Poincar\'e SUGRA 
has an inconsistency pointed out in Ref.~\cite{Hall:2003yc}. 
This fact can be seen as an explicit mass term for the 
graviphoton in our framework. 

We can go back to the ordinary gauge fixing (\ref{eq:utuf}) 
by rotating ${\cal A}^a_{\ i}$ for the index $a$. 
Then, the following additional terms~\cite{Abe:2005wn} come out, 
\begin{eqnarray}
e^{-1}{\cal L}_\omega &=& 
f'(y) (i\vec\omega \cdot \vec\sigma)_{ab} 
\bigg\{ \epsilon^{ij} 
({\cal A}^a_{\ j}\nabla_4{\cal A}^b_{\ i} 
-{\cal A}^b_{\ i}\nabla_4{\cal A}^a_{\ j}) 
+2i\bar\zeta^b \gamma^4 \zeta^a 
\nonumber \\ &&
-4i\bar\psi_m^i \gamma^4 \gamma^m \zeta^a {\cal A}^b_{\ i}
+2i\bar\psi_m^{(i} \gamma^{m4n}\psi_n^{j)} 
{\cal A}^b_{\ j}{\cal A}^a_{\ i} 
-\frac{1}{\cal N}\epsilon^{jk}{V_4}^i_{\ k} 
({\cal A}^b_{\ i}{\cal A}^a_{\ j} 
+{\cal A}^b_{\ j}{\cal A}^a_{\ i}) \bigg\} 
\nonumber \\ &&
-(f'(y)|\vec\omega|)^2 \epsilon^{ij}\epsilon_{ab} 
{\cal A}^b_{\ i}{\cal A}^a_{\ j}, 
\end{eqnarray}
in the bulk Lagrangian, where $f'(y)=(d/dy)f(y)$. 
In this basis, the total Lagrangian is given by 
\begin{eqnarray}
{\cal L} 
&=& {\cal L}_b + {\cal L}_f 
+{\cal L}_{\rm tension} 
+{\cal L}_w +{\cal L}_\omega 
+{\cal L}_{\rm aux} 
\big|_{{\cal F}^a_{\ i} \to {\cal F}^a_{\ i}+C^a_{\ i}}, 
\label{eq:twistedlagrangian}
\end{eqnarray}
where the compensator is now in the periodic basis 
which follows the normal gauge fixing (\ref{eq:scgf}).

\subsection{On-shell Lagrangian}
\label{ssec:oslag}
The terms containing ${V_4}^{ij}$ in the Lagrangian 
are collected as 
\begin{eqnarray}
e^{-1}{\cal L}_{{V_4}^{ij}} &=& 
\epsilon^{jk} \epsilon_{li} 
(V_4-V_{4\,{\rm sol}})^i_{\ k} 
(V_4-V_{4\,{\rm sol}})^l_{\ j} 
+2(A+B)^j_{\ i} {V_4}^i_{\ j}, 
\end{eqnarray}
where 
\begin{eqnarray}
A^i_{\ j} &=& 
-\frac{1}{{\cal N}}f'(y) \epsilon^{ik} \big\{ 
(\omega_2-i\omega_1){\cal A}^{a=1}_{\ k}{\cal A}^{a=1}_{\ j}
+(\omega_2+i\omega_1){\cal A}^{a=2}_{\ k}{\cal A}^{a=2}_{\ j} \big\}, 
\\
B^i_{\ j} &=& 
\left( \begin{array}{cc}
{\cal A}^{a=2}_{\ i=1}B & {\cal A}^{a=2}_{\ i=2}B \\
-({\cal A}^{a=2}_{\ i=2}B)^\ast & ({\cal A}^{a=2}_{\ i=1}B)^\ast 
\end{array} \right), 
\label{eq:defAB}
\end{eqnarray}
and 
\begin{eqnarray}
B &=& e^{-1}e_{(4)} \sum_{l=0,\pi} 
\delta(y-lR)\,(w_2^{(l)}+iw_1^{(l)}){\cal A}^{a=2}_{\ i=2}. 
\end{eqnarray}
Here we have used the relations 
$V^{i=1}_{\ j=1}=(V^{i=2}_{\ j=2})^\ast$, 
$V^{i=1}_{\ j=2}=-(V^{i=2}_{\ j=1})^\ast$, 
and 
${\cal A}^{a=1}_{\ i=1}=({\cal A}^{a=2}_{\ i=2})^\ast$, 
${\cal A}^{a=1}_{\ i=2}=-({\cal A}^{a=2}_{\ i=1})^\ast$. 

Then we easily find 
\begin{eqnarray}
e^{-1}{\cal L}_{{V_4}^{ij}} &=& 
-\epsilon^{jk} \epsilon_{li} \bigg\{
-\Big( V_4-(V_{4\,{\rm sol}}+A+B) \Big)^i_{\ k} 
\Big( V_4-(V_{4\,{\rm sol}}+A+B) \Big)^j_{\ l} 
\nonumber \\ && \qquad\qquad 
+2(A+B)^i_{\ k} {V_{4\,{\rm sol}}}^l_{\ j} 
+(A+B)^i_{\ k} (A+B)^l_{\ j} \bigg\}, 
\label{eq:vlag}
\end{eqnarray}
and the total Lagrangian (\ref{eq:twistedlagrangian}) 
can be rewritten as 
\begin{eqnarray}
{\cal L} 
&=& {\cal L}_b + {\cal L}_f 
+{\cal L}'_{\rm tension} 
+{\cal L}'_{w+\omega} 
+{\cal L}_{\rm aux} 
\Bigg|_{\mbox{\scriptsize $
\begin{array}{l}
{\cal F}^a_{\ i} \to {\cal F}^a_{\ i}+C^a_{\ i} \\
{V_4}^i_{\ j} \to {V_4}^i_{\ j}-(A+B)^i_{\ j} 
\end{array}$}}, 
\end{eqnarray}
where 
\begin{eqnarray}
e^{-1}{\cal L}'_{\rm tension} &=& 
-4g_R M^{I=0} \sum_{l=0,\pi} \delta^l(y) 
\big( q_1 w^{(l)}_2 - q_2 w^{(l)}_1 \big) 
\Big( 1+{\cal A}_{\underline\alpha}^{\ i} 
{\cal A}^{\underline\alpha}_{\ i} \Big) 
\nonumber \\ && \qquad 
+2f'(y) \sum_{l=0,\pi} \delta^l(y) 
(\omega_2 w^{(l)}_2+\omega_1 w^{(l)}_1) 
\Big( 1+{\cal A}_{\underline\alpha}^{\ i} 
{\cal A}^{\underline\alpha}_{\ i} \Big)^2, 
\\
e^{-1}{\cal L}'_{w+\omega} &=& 
\bigg\{ 
\sum_{l=0,\pi} \delta^l(y) 
(i\vec{w}^{(l)} \cdot \vec\sigma)_{ij} 
+f'(y) (i\vec\omega \cdot \vec\sigma)_{ij} \bigg\} 
\nonumber \\ && \times 
\bigg\{ 
2i\bar\zeta^{\underline\beta} 
\gamma^4 \zeta^{\underline\alpha} 
{\cal A}_{\underline\beta}^{\ j} 
{\cal A}_{\underline\alpha}^{\ i} 
\Big( 1+{\cal A}_{\underline\gamma}^{\ k} 
{\cal A}^{\underline\gamma}_{\ k} \Big)^{-1} 
+4i\bar\psi_m^i \gamma^4 \gamma^m 
\zeta^{\underline\alpha} {\cal A}_{\underline\alpha}^{\ j} 
\nonumber \\ && \qquad 
+\Big(2i\bar\psi_m^{(i} \gamma^{m4n}\psi_n^{j)}
-2{\cal A}^{\underline\alpha (i} \nabla_4 
{\cal A}_{\underline\alpha}^{\ j)} 
+i{\cal N}_{IJ} \bar\Omega^{Ii} \gamma_4 \Omega^{Jj} \Big) 
\Big( 1+{\cal A}_{\underline\beta}^{\ k} 
{\cal A}^{\underline\beta}_{\ k} \Big) \bigg\} 
\nonumber \\ &&
-2\bigg\{ \sum_{l=0,\pi} 
\Big( \delta^l(y) |\vec{w}^{(l)}| \Big)^2
+\Big( f'(y)|\vec\omega| \Big)^2 \bigg\} 
\Big( 1+{\cal A}_{\underline\alpha}^{\ i} 
{\cal A}^{\underline\alpha}_{\ i} \Big) 
{\cal A}_{\underline\beta}^{\ j} 
{\cal A}^{\underline\beta}_{\ j}, 
\end{eqnarray}
and 
\begin{eqnarray}
\delta^l(y) &=& e^{-1}e_{(4)} \delta(y-lR). 
\end{eqnarray}
Note that the above expression is valid 
under the regularization satisfying 
$\delta^l(y) F(\Phi_+,\Phi_-) = \delta^l(y) F(\Phi_+,0)$ 
for any regular function $F$ of any $Z_2$-even 
and $Z_2$-odd fields $\Phi_+$ and $\Phi_-$. 

Taking into account the superconformal gauge fixing conditions 
(\ref{eq:scgf}) which result in 
$M^{I=0}=1+\cdots$ and ${\cal A}^a_{\ i}=\delta^a_{\ i}+\cdots$, 
the tensions at the boundaries are written as 
\begin{eqnarray}
{\cal L}'_{\rm tension} &=& 
-e_{(4)} \sum_{l=0,\pi} \delta(y-lR) \Lambda^{(l)} 
+\cdots, 
\label{eq:wtwistbt} \\
\Lambda^{(l)} &=& 
2 \Big( 2g_R q_1-f'(lR) \omega_2 \Big) w^{(l)}_2 
-2 \Big( 2g_R q_2+f'(lR) \omega_1 \Big) w^{(l)}_1, 
\end{eqnarray}
where ellipses stand for field-dependent terms. 
The bulk cosmological constant $\Lambda$ induced by the 
$U(1)_R$ gauging is found in ${\cal L}_b$ as 
\begin{eqnarray}
{\cal L}_b &\supset& 
{\cal L}_{\rm c.c.} \ = \ 
\frac{8}{3} g_R^2 (M^{I=0})^2 |{\cal A}^{a=2}_{\ i=2}|^2 
\ = \ -\Lambda + \cdots, 
\label{eq:cc}
\end{eqnarray}
where $\Lambda=-6k^2$ and $k=2g_R/3$ is the AdS$_5$ 
curvature scale in Eq.~(\ref{eq:ads5}). 
The ellipsis again stands for field dependent terms. 
By comparing (\ref{eq:wtwistbt}) and (\ref{eq:cc}), 
we find that the RS relation (\ref{eq:rs}) is satisfied if 
\begin{eqnarray}
6k &=& 
-\Big( -6kq_1 +2f'(0) \omega_2 \Big) w^{(0)}_2
-\Big( 6kq_2 +2f'(0) \omega_1 \Big) w^{(0)}_1 
\nonumber \\ &=& 
\Big( -6kq_1 +2f'(\pi R) \omega_2 \Big) w^{(\pi)}_2
+\Big( 6kq_2 +2f'(\pi R) \omega_1 \Big) w^{(\pi)}_1. 
\label{eq:rsrelation0}
\end{eqnarray}

Together with the consistency condition (\ref{eq:ccfrg}) 
for the twist vector $\vec\omega$, the relation 
(\ref{eq:rsrelation0}) results in 
\begin{eqnarray}
1 &=& 
(w^{(0)}_2 q_1 
-w^{(0)}_1 q_2) 
-\frac{\omega}{3k}f'(0) (w^{(0)}_1 q_1 + w^{(0)}_2 q_2) 
\nonumber \\ &=& 
-(w^{(\pi)}_2 q_1 
-w^{(\pi)}_1 q_2) 
+\frac{\omega}{3k}f'(\pi R) (w^{(\pi)}_1 q_1 + w^{(\pi)}_2 q_2). 
\label{eq:rsrelation}
\end{eqnarray}
For instance, if the $U(1)_R$ gauging direction 
is $\vec{q}=(1,\,0,\,0)$ and the boundary constant 
superpotentials $W^{(0,\pi)}$ are real, 
the RS condition (\ref{eq:rsrelation})
uniquely determines 
$W^{(0,\pi)}=w_2^{(0,\pi)}$ as 
\begin{eqnarray}
w_2^{(0)} &=& -w_2^{(\pi)} \ = \ 1. 
\qquad (q_2=0,\,w_1^{(0,\pi)}=0) 
\label{eq:simpleex}
\end{eqnarray}

\section{Killing spinor equations and SUSY breaking}
\label{sec:killing}
Now we study the Killing conditions. 
The gravitino SUSY transformation is given by 
\begin{eqnarray}
\delta \psi_\mu^i &=& 
\bigg( \partial_\mu 
-\frac{1}{4} \omega_\mu^{\ mn} \gamma_{mn} 
+\frac{1}{2}b_\mu \bigg) \varepsilon^i 
-{V_\mu}^i_{\ j} \varepsilon^j 
+\frac{1}{2}v^{mn} \gamma_{\mu mn} \varepsilon^i 
-\gamma_\mu \eta^i, 
\label{eq:gravitinost} \\
\eta^i &=& 
-\frac{{\cal N}_I}{12{\cal N}} 
\gamma \cdot \hat{F}^I(W) \varepsilon 
+\frac{{\cal N}_I}{3{\cal N}} 
{Y^I}^i_{\ j} \varepsilon^j  
+\frac{{\cal N}_{IJ}}{3{\cal N}} 
{\Omega^{I}}^i 2i \bar\varepsilon \Omega^J, 
\end{eqnarray}
where $\varepsilon^i$ is SUSY transformation parameter. 
To find a Killing spinor in the AdS$_5$ background (\ref{eq:ads5}), 
we substitute the on-shell values of the auxiliary fields 
into Eq.~(\ref{eq:gravitinost}), and apply the superconformal 
gauge fixing conditions (\ref{eq:scgf}). 

From $\delta \psi_{\underline\mu}^i=0$, we find 
\begin{eqnarray}
\frac{1}{3}{\cal N}_I{Y^I}^i_{\ j}\varepsilon^j 
&=& -\frac{1}{2} \sigma'(y) \gamma_4 \varepsilon^i, 
\label{eq:vnsdpsimu}
\end{eqnarray}
where $\sigma'(y)=(d/dy)\sigma(y)=k \epsilon(y)$. 
This results in 
\begin{eqnarray}
\sigma'(y) \varepsilon^{i=1} 
-\frac{2}{3}g_R |{\cal A}^{a=2}_{\ i=2}|^2 M^{I=0} 
(q_1-iq_2) \gamma_5 \varepsilon^{i=2} &=& 0, 
\nonumber \\
\sigma'(y) \varepsilon^{i=2} 
-\frac{2}{3}g_R |{\cal A}^{a=2}_{\ i=2}|^2 M^{I=0} 
(q_1+iq_2) \gamma_5 \varepsilon^{i=1} &=& 0. 
\label{eq:ke1}
\end{eqnarray}
Substituting these equations into another 
condition $\delta \psi_y=0$ yields 
\begin{eqnarray}
\partial_y \varepsilon^i 
-{V_y}^i_{\ j} \varepsilon^j 
-\frac{1}{2} \sigma'(y) \varepsilon^i 
&=& 0. 
\label{eq:kscdn1}
\end{eqnarray}

The transformation parameter $\varepsilon^i$ is an 
$SU(2)$-Majorana spinor, and is recasted as two Majorana spinors 
$\varepsilon_+=\varepsilon^{i=1}_R+\varepsilon^{i=2}_L$ and 
$\varepsilon_-=i(\varepsilon^{i=1}_L+\varepsilon^{i=2}_R)$, 
which are $Z_2$-even and $Z_2$-odd, respectively. 
Then we obtain a relation between $\varepsilon_+$ and 
$\varepsilon_-$ from Eq.~(\ref{eq:ke1}) as 
\begin{eqnarray}
\varepsilon_- &=& -(q_2-iq_1 \gamma_5)\epsilon(y) \varepsilon_+. 
\label{eq:emep}
\end{eqnarray}
From (\ref{eq:kscdn1}) and (\ref{eq:emep}), we obtain 
\begin{eqnarray}
\left\{ 
\begin{array}{lcl}
\partial_y \tilde\varepsilon_{+R} 
&=& -i(q_2-q_1)({V_y}^{i=2}_{\ j=1} )^\ast 
\epsilon(y) \tilde\varepsilon_{+R}, \\*[5pt]
\partial_y \tilde\varepsilon_{+L} 
&=& i(q_2+iq_1){V_y}^{i=2}_{\ j=1} 
\epsilon(y) \tilde\varepsilon_{+L}, \\*[5pt]
\partial_y \big( \epsilon(y) \tilde\varepsilon_{+R} \big) 
&=& -i(q_2+iq_1){V_y}^{i=2}_{\ j=1} 
\tilde\varepsilon_{+R}, \\*[5pt]
\partial_y \big( \epsilon(y) \tilde\varepsilon_{+L} \big) 
&=& i(q_2-iq_1)({V_y}^{i=2}_{\ j=1})^\ast \tilde\varepsilon_{+L}, 
\end{array}
\right. 
\label{eq:vnsdpsiy}
\end{eqnarray}
where 
$\tilde\varepsilon_\pm \equiv e^{-\sigma(y)/2} \varepsilon_\pm$. 

To have a Killing spinor, ${V_y}^{i=2}_{\ j=1}$ must be 
in the form of 
\begin{eqnarray}
{V_y}^{i=2}_{\ j=1} &=& 
(q_1+iq_2) \big( -iv+\partial_y \epsilon(y) \big), 
\label{eq:susyvy}
\end{eqnarray}
where $v$ is a real constant. 
The on-shell form of $V_y=e_y^{\ 4}V_4$ is determined 
by Eq.~(\ref{eq:vlag}) and we find 
\begin{eqnarray}
{V_y}^{i=2}_{\ j=1} &=& (A+B)^{i=2}_{\ j=1}, 
\label{eq:onshellvy}
\end{eqnarray}
where $A^i_{\ j}$ and $B^i_{\ j}$ are given in Eq.~(\ref{eq:defAB}). 
Comparing Eq.~(\ref{eq:onshellvy}) with Eq.~(\ref{eq:susyvy}), 
we find that, if $f(y)$ is not a singular function, 
\begin{eqnarray}
w^{(0)}_1=2q_2, \quad 
w^{(0)}_2=-2q_1, \quad 
w^{(\pi)}_1=-2q_2, \quad 
w^{(\pi)}_2=2q_1, 
\label{eq:bpsw}
\end{eqnarray}
is the condition for preserved SUSY in a slice of AdS$_5$, 
which leads to relations, 
\begin{eqnarray}
q_2 w^{(0)}_1 - q_1 w^{(0)}_2 
&=& -(q_2 w^{(\pi)}_1-q_1 w^{(\pi)}_2) 
\ = \ 2(q_1^2+q_2^2) \ = \ 2, 
\end{eqnarray}
and 
\begin{eqnarray}
w^{(0)}_1 q_1 + w^{(0)}_2 q_2 
\ = \ w^{(\pi)}_1 q_1 + w^{(\pi)}_2 q_2 
&=& 0. 
\end{eqnarray}
From these relations, we find that the RS relation 
(\ref{eq:rsrelation}) is never satisfied for SUSY-preserving 
parameter choice (\ref{eq:bpsw}), 
{\it irrespective of} the values of twist parameter $\omega$. 
Note that SUSY might be restored in a detuned case with an 
AdS$_4$ background geometry.

\section{4D gaugino and gravitino masses}
\label{sec:masses}
From the analysis of Killing spinor, we have found that 
SUSY is spontaneously broken in the orbifold slice of AdS$_5$. 
The result also shows that the SS twist cannot restore SUSY 
with vanishing 4D cosmological constant. 
This is contrast to the situation in the case of $Z_2$-odd 
coupling, where the tensions at the boundaries are 
automatically supplied by the four-form mechanism generating 
such an odd coupling constant, and SUSY can be preserved 
in an orbifold slice of AdS$_5$. 

Because SUSY must be broken with almost vanishing vacuum energy 
in our real world, the model studied above can be one of 
interesting mechanisms for such realization. Then, next, we study 
SUSY breaking structure more precisely by analyzing the 4D 
superparticle spectrum. For simplicity, we consider the case with 
a physical Abelian vector multiplet ($n_V=1$) with $Z_2$-even parity 
and no physical hypermultiplet ($n_H=0$) coupled to SUGRA, and show the 
lightest masses in the gaugino and gravitino spectra. 

The gaugino and gravitino bilinear terms in our model are collected as 
\begin{eqnarray}
e^{-1}{\cal L}^{(2)}_{1/2} 
&=& 2ia_{IJ} \bar\Omega^{Ii} \nabla \!\!\!\!/\,\, \Omega^J_{\ i} 
+2i{\cal N}^{IJ}{\cal N}_{JKL} {\cal A}_{ai} (gt_I)^{ab} {\cal A}_{bj} 
\bar\Omega^{Ki} \Omega^{Lj} 
+2ia_{IJ} ({\cal A}^a_{\ i} \partial_y {\cal A}_{aj}) 
\bar\Omega^{Ii} \gamma^y \Omega^{Jj} 
\nonumber \\ &=& 
i \bar\Omega^i \nabla \!\!\!\!/\,\, \Omega_{\ i} 
-i\bar\Omega^i \bigg( \frac{1}{2}m^{(q)}_{ij}
-i\gamma_5 m^{(\omega)}_{ij} \bigg) \Omega^j 
+\cdots, 
\\
e^{-1}{\cal L}^{(2)}_{3/2} 
&=& -2i \bar\psi^i_{\mu} \gamma^{\mu \nu \rho} \nabla_{\nu} \psi_{\rho i} 
-2i M^I {\cal A}^a_{\ i} (gt_I)_{ab} {\cal A}^b_{\ j} 
\bar\psi^{(i}_m \gamma^{mn} \psi^{j)}_n 
-2i {\cal A}^a_{\ i} \nabla_n {\cal A}_{aj}
\bar\psi^{(i}_m \gamma^{mnk} \psi^{j)}_k 
\nonumber \\ &=& 
2i \bar\psi^i_{\mu} \gamma^{\mu \nu \rho} \nabla_{\rho} \psi_{\nu i} 
+2i \bar\psi^{(i}_m \gamma^{mn} 
\bigg( \frac{3}{2}m^{(q)}_{ij}
-i\gamma_5 m^{(\omega)}_{ij} \bigg) \psi^{j)}_n 
+\cdots, 
\label{eq:bilinearterms}
\end{eqnarray}
where the compensator scalar ${\cal A}^a_{\ i}$ 
is in the twisted basis (\ref{eq:tuf}). 
The ellipses denote cubic or higher-order terms, 
and the mass parameters are given by 
\begin{eqnarray}
m^{(q)}_{ij} &=& k 
\left( \begin{array}{cc}
q_2-iq_1 & 0 \\
0 & q_2+iq_1 
\end{array} \right), 
\\
m^{(\omega)}_{ij} &=& 
f'(y)
\left( \begin{array}{cc}
\omega_2-i\omega_1 & 0 \\
0 & \omega_2+i\omega_1 
\end{array} \right) 
\nonumber \\ && \qquad 
-\sum_{l=0,\pi}e^{-1}e_{(4)} \delta (y-lR) 
\left( \begin{array}{cc}
w^{(l)}_2-iw^{(l)}_1 & 0 \\
0 & w^{(l)}_2+iw^{(l)}_1 
\end{array} \right). 
\end{eqnarray}

Both $\Omega^i$ and $\psi^i_\mu$ are $SU(2)$-Majorana spinors 
satisfying $\bar\Omega^i \equiv 
(\Omega_i)^\dagger \gamma^0 = (\Omega^i)^T C_5$ 
where $C_5$ is a 5D charge conjugation matrix. We can decompose 
them into two Weyl spinors as, e.g., 
\begin{eqnarray}
\Omega^{i=1} &=& 
\left( \begin{array}{c}
\lambda_\alpha \\ -i\bar\xi^{\dot\alpha} 
\end{array} \right), \qquad 
\Omega^{i=2} \ = \ 
\left( \begin{array}{c}
-i\xi_\alpha \\ \bar\lambda^{\dot\alpha} 
\end{array} \right). 
\end{eqnarray}
We rearrange them into usual Majorana spinors $\Omega_\pm$ 
in the $Z_2$-eigenstates satisfying 
$(\Omega_\pm)^\dagger \gamma^0 = (\Omega_\pm)^T C_4$, 
where $C_4$ is a 4D charge conjugation matrix, as
\begin{eqnarray}
\Omega_+ &=& 
\left( \begin{array}{c}
\lambda_\alpha \\ \bar\lambda^{\dot\alpha} 
\end{array} \right), \qquad 
\Omega_- \ = \ 
\left( \begin{array}{c}
\xi_\alpha \\ \bar\xi^{\dot\alpha} 
\end{array} \right). 
\end{eqnarray}

These spinors are expanded into Kaluza-Klein (KK) modes as 
\begin{eqnarray}
\lambda_\alpha 
&=& e^{2\sigma(y)} \sum_n f^{(+)}_n(y) \lambda^{(n)}_\alpha(x), 
\nonumber \\
\xi_\alpha 
&=& e^{2\sigma(y)} \sum_n f^{(-)}_n(y) \xi^{(n)}_\alpha(x), 
\end{eqnarray}
where complex wavefunctions $f^{(\pm)}_n(y)$ are 
solutions to the following equations, 
\begin{eqnarray}
i\partial_y f^{(-)}_n(y)
+e^{\sigma(y)}M_n (f^{(+)}_n(y))^\ast 
+(\mu^{(1)}_+ + i\mu^{(2)}_+) f^{(+)}_n(y) &=& 0, 
\nonumber \\
i\partial_y f^{(+)}_n(y)
-e^{\sigma(y)}M_n (f^{(-)}_n(y))^\ast 
-(\mu^{(1)}_- + i\mu^{(2)}_-) f^{(-)}_n(y) &=& 0. 
\label{eq:kkeq}
\end{eqnarray}
Here, $M_n$ is the KK mass eigenvalues, and 
\begin{eqnarray}
\mu^{(1)}_\pm &=& skq_1 \pm {\rm Re}\,m^{(\omega)}_{22}, 
\nonumber \\
\mu^{(2)}_\pm &=& \pm skq_2 - {\rm Im}\,m^{(\omega)}_{22}, 
\label{eq:mu12}
\end{eqnarray}
where $s=1/2,\,3/2$ for gaugino and gravitino, respectively. 

For a simple parameter choice (\ref{eq:simpleex}) 
and a gauge fixing function $f(y)=y/R$, 
the KK equations (\ref{eq:kkeq}) are written as 
\begin{eqnarray}
\partial_y a^{(+)}_n(y)+(e^{\sigma(y)}M_n-sk)a^{(-)}_n(y)
-(\omega/R)\,b^{(-)}_n(y) &=& 0, 
\nonumber \\
\partial_y a^{(-)}_n(y)-(e^{\sigma(y)}M_n+sk)a^{(+)}_n(y)
-(\omega/R)\,b^{(+)}_n(y) &=& 0, 
\nonumber \\
\partial_y b^{(+)}_n(y)-(e^{\sigma(y)}M_n+sk)b^{(-)}_n(y)
+(\omega/R)\,a^{(-)}_n(y) &=& 0, 
\nonumber \\
\partial_y b^{(-)}_n(y)+(e^{\sigma(y)}M_n-sk)b^{(+)}_n(y)
+(\omega/R)\,a^{(+)}_n(y) &=& 0, 
\label{eq:kkeqreal}
\end{eqnarray}
where real wavefunctions $a^{(\pm)}_n(y)$ and $b^{(\pm)}_n(y)$ 
are defined by 
\begin{eqnarray}
f^{(+)}_n(y) &=& a^{(+)}_n(y)+ib^{(+)}_n(y), 
\nonumber \\
f^{(-)}_n(y) &=& i(a^{(-)}_n(y)+ib^{(-)}_n(y)). 
\end{eqnarray}
The boundary conditions can be extracted from the 
delta-functions in $m^{(\omega)}_{22}$ of Eq.~(\ref{eq:mu12}), 
and are found in this case as 
\begin{eqnarray}
2a^{(-)}_n(0+\varepsilon)+a^{(+)}_n(0) &=& 0, 
\nonumber \\
2a^{(-)}_n(\pi R-\varepsilon)+a^{(+)}_n(\pi R) &=& 0, 
\nonumber \\
2b^{(-)}_n(0+\varepsilon)+b^{(+)}_n(0) &=& 0, 
\nonumber \\
2b^{(-)}_n(\pi R-\varepsilon)+b^{(+)}_n(\pi R) &=& 0, 
\end{eqnarray}
where $\varepsilon$ is a positive infinitesimal. 
Note that $a^{(-)}(y)$ and $b^{(-)}(y)$ are 
discontinuous at the boundaries. 

\begin{figure}[t]
\begin{center}
\epsfig{file=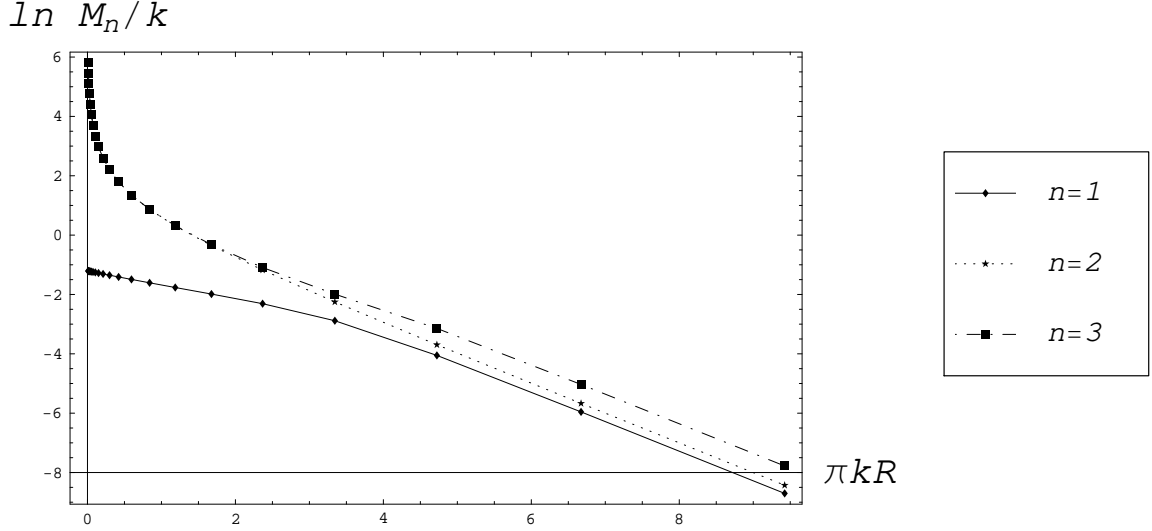,width=\linewidth} \\
(a) Gaugino: $s=1/2$,\, $\omega=0$ \\
\epsfig{file=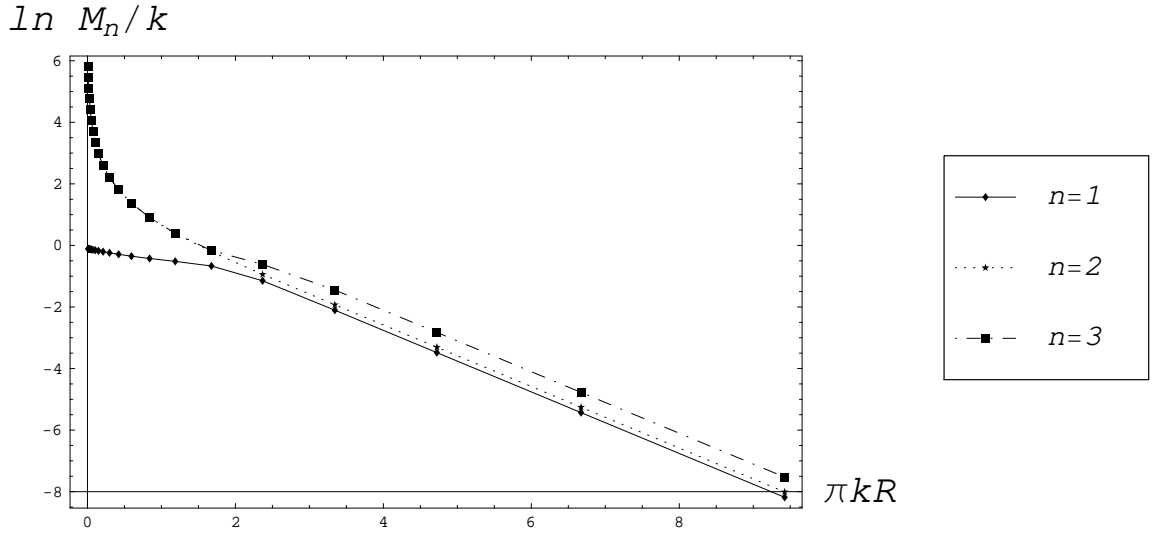,width=\linewidth} \\
(b) Gravitino:  $s=3/2$,\, $\omega=0$ 
\end{center}
\caption{Numerical results for (a) the gaugino and 
(b) the gravitino mass spectra up to the third KK 
excited modes in terms of a dimensionless quantity 
$\pi k R$. The parameters in the model are chosen 
as in Eq.~(\ref{eq:simpleex}).}
\label{fig:3rdexmd}
\end{figure}

\begin{figure}[t]
\begin{center}
\epsfig{file=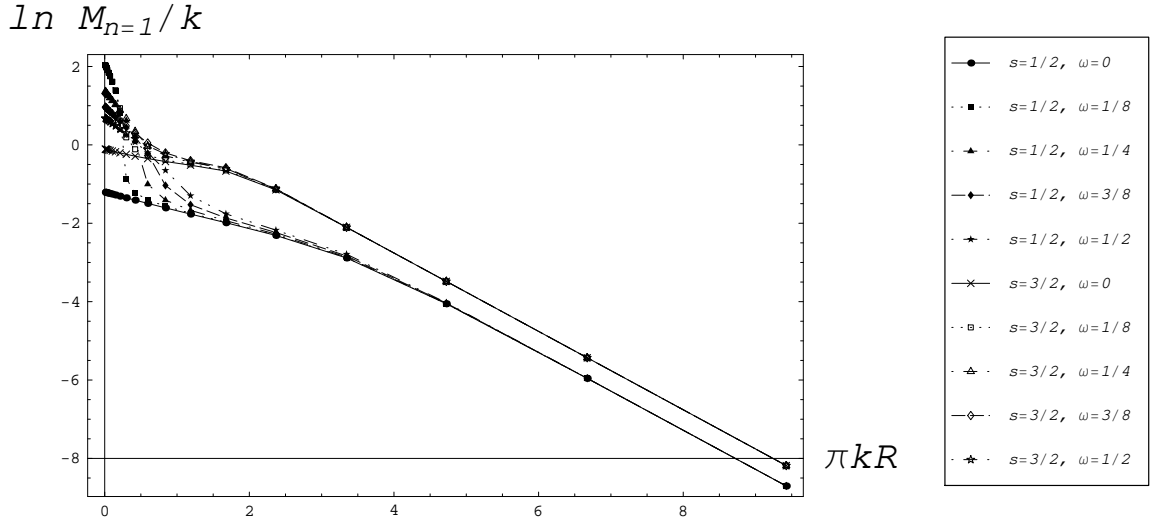,width=\linewidth} \\
(a) The lightest gaugino ($s=1/2$) and gravitino ($s=3/2$) masses. 
\epsfig{file=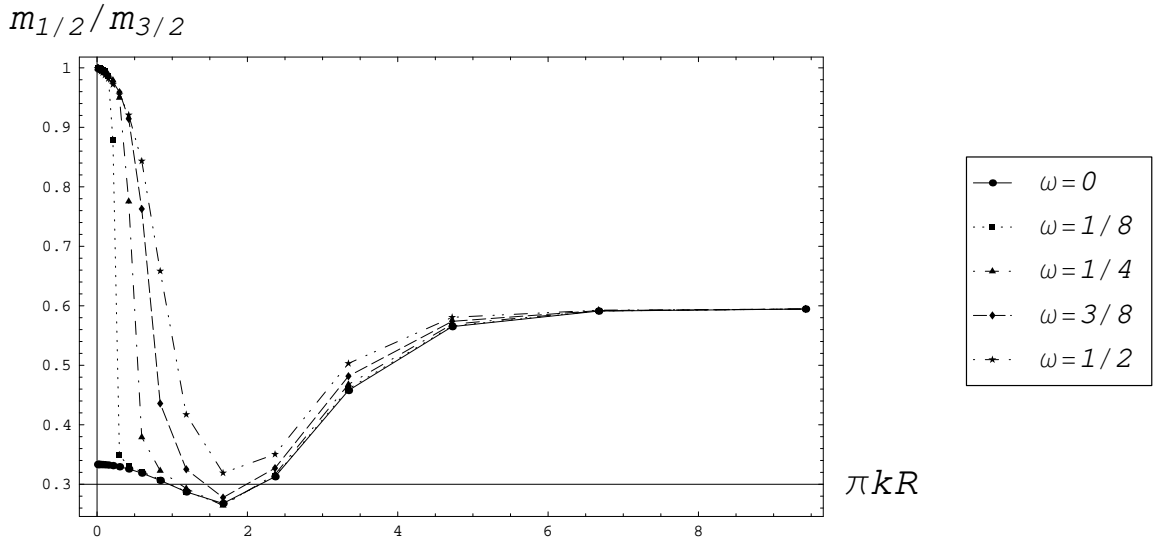,width=\linewidth} \\
(b) The lightest gaugino/gravitino mass ratio. 
\end{center}
\caption{Numerical estimation for the effect of the SS twist 
$\omega$ on the masses of the lightest modes in the gaugino 
and the gravitino KK spectra. (a) The lightest gaugino and 
gravitino masses are shown together, and (b) the lightest 
gaugino/gravitino mass ratio is shown for various values 
of the SS twist, $\omega$. The parameters in the model are 
chosen as in Eq.~(\ref{eq:simpleex}).}
\label{fig:lightestmd}
\end{figure}

\begin{figure}[t]
\begin{center}
\epsfig{file=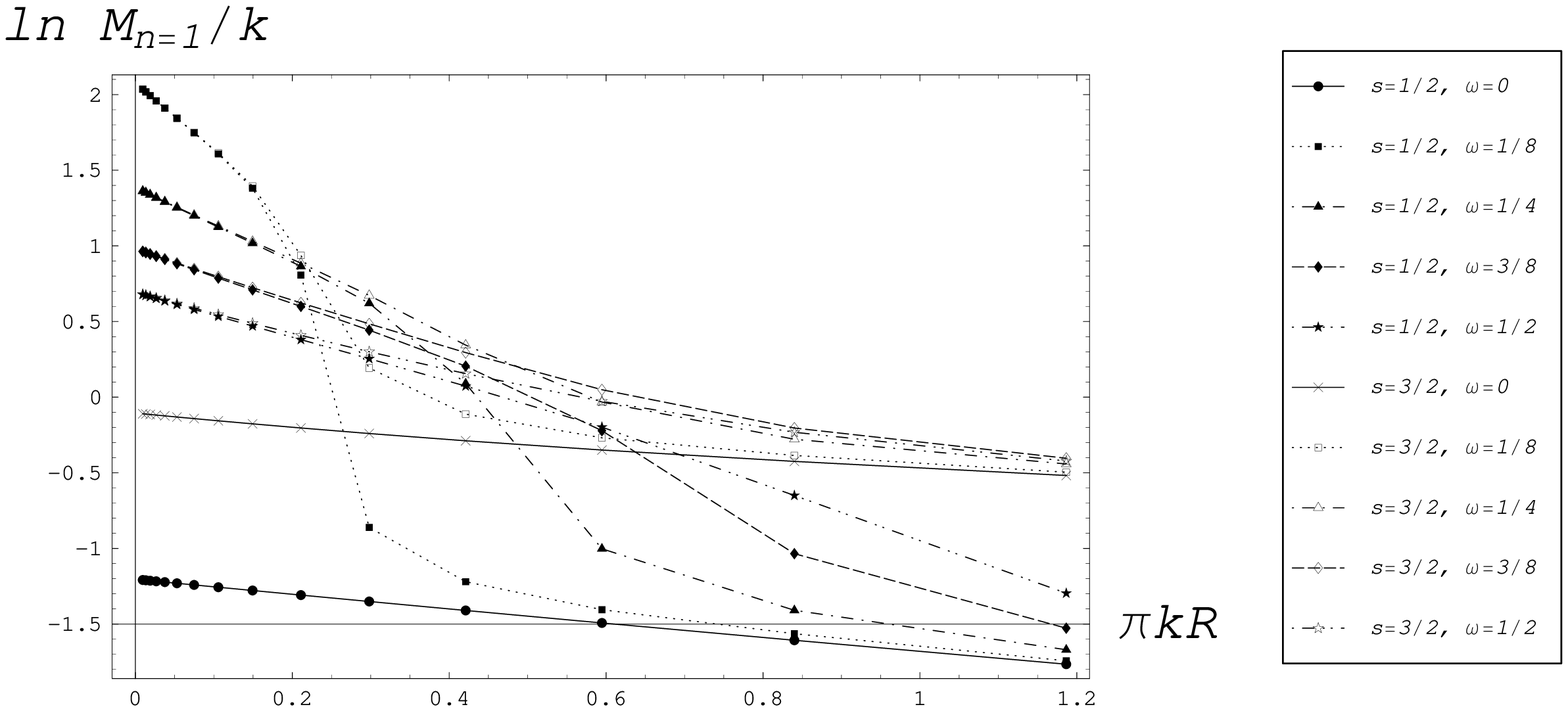,width=\linewidth} \\
(a) The lightest gaugino ($s=1/2$) and gravitino ($s=3/2$) masses. 
\epsfig{file=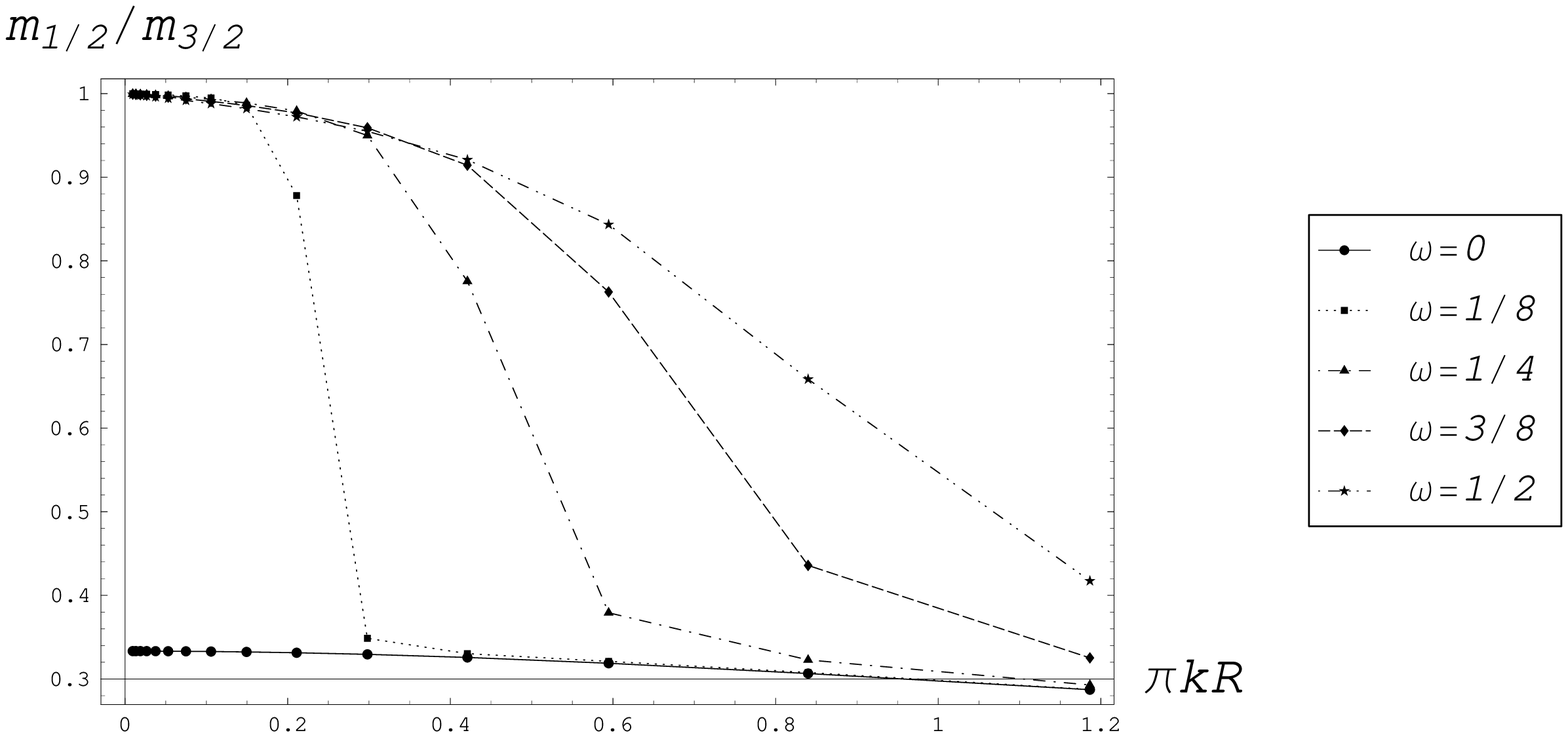,width=\linewidth} \\
(b) The lightest gaugino/gravitino mass ratio. 
\end{center}
\caption{The small $\pi k R$ region of 
Fig.~\ref{fig:lightestmd} is magnified.}
\label{fig:lightestmdlg}
\end{figure}

In the case of SUGRA with $Z_2$-odd coupling, 
Dirac-type bulk mass terms such as $\bar\Omega_+ \Omega_-$ 
are generated, which results in the Bessel-type wavefunctions. 
On the other hand, in the case of $Z_2$-even coupling, 
we have Majorana-type mass terms such as 
$\bar\Omega_+ \Omega_+$, $\bar\Omega_- \Omega_-$. 
In the latter case, we cannot solve analytically the 
first-order coupled differential equations (\ref{eq:kkeqreal}). 
Therefore, we show some numerical results for the 
mass eigenvalues in the case of simple parameter choice 
shown in Eq.~(\ref{eq:simpleex}). 

In Fig.~\ref{fig:3rdexmd}, the $\pi k R$ dependences of 
the gaugino and gravitino mass spectra are given for 
$\omega=0$ up to the third KK excited modes. 
In Fig.~\ref{fig:lightestmd} (a), the masses of the lightest 
modes are shown together for $\omega=0,\,1/8,\,1/4,\,3/8,\,1/2$. 
The small $\pi k R$ region of Fig.~\ref{fig:lightestmd} (a) 
is magnified in Fig.~\ref{fig:lightestmdlg} (a). 
From these figures, we find that, for $\pi k R \gg 1$, 
the lightest gaugino and gravitino masses behave like 
$m_{1/2},\,m_{3/2} \approx e^{-\pi k R}k$ 
which is exponentially suppressed as is expected from the 
highly warped geometry. The effect of the SS twist is 
negligible in this region. On the other and, 
it becomes more important in the region $\pi k R < 2$. 

The overall SUSY breaking scale can be read off from the 
lightest gravitino mass $m_{3/2}$. The next important quantity 
is the gaugino/gravitino mass ratio $m_{1/2}/m_{3/2}$, which 
is shown in Fig.~\ref{fig:lightestmd} (b). 
The small $\pi k R$ region of Fig.~\ref{fig:lightestmd} (b) 
is magnified in Fig.~\ref{fig:lightestmdlg} (b). 
The ratio takes minimum value at $\pi k R \sim 2$ which 
is about $0.2$. 
In the flat limit $k \to 0$, the ratio $m_{1/2}/m_{3/2}$ 
converges on unity when $\omega \ne 0$. This is the same 
ratio as in the case of the radion mediation in flat space 
which is equivalent to the SS SUSY breaking in flat 
space~\cite{vonGersdorff:2001ak}. When $\omega=0$, 
SUSY is restored in the flat limit, and both 
$m_{1/2}$ and $m_{3/2}$ approach to zero.

\section{Summary and discussions}
\label{sec:conclusion}
We have studied the five-dimensional SUGRA compactified 
on an orbifold $S^1/Z_2$ where the $U(1)_R$ symmetry is gauged by 
the graviphoton with $Z_2$-even coupling. In order to include 
superpotential terms at the orbifold boundaries as well as 
matter fields, we worked in the off-shell formulation of 5D SUGRA. 
For generality, we have introduced the SS twist parameter in the 
Lagrangian. After integrating out the auxiliary fields, the on-shell 
action is obtained including tensions induced by the constant 
superpotentials at the boundaries, and the Killing spinor equations 
are derived in an orbifold slice of AdS$_5$. 
It has been shown that the equations do not allow a Killing spinor 
on such background {\it irrespective of} the value of twist parameter 
$\omega$, that means SUSY is spontaneously broken on this 
background and the SS twist cannot restore it.\footnote{
We have to remark that, because SUSY is broken, 
our model does not correspond to an off-shell 
construction of Altendorfer-Bagger-Nemeschansky (ABN) 
model~\cite{Altendorfer:2000rr}. 
We may need another setup and/or a nontrivial regularization 
of the orbifold singularities for such construction.} 

We have also shown the details of SUSY breaking 
when the constant superpotential terms at the boundaries are 
tuned such that the 4D cosmological constant vanishes. 
The KK equations for the bulk gaugino and 
gravitino wavefunctions are derived, which do not have an 
analytic expressions for the solutions in contrast to the case of 
SUGRA with $Z_2$-odd coupling. This is due to the existence 
of the Majorana-type bulk mass instead of the Dirac-type. 
A numerical computation for the KK mass eigenvalues has 
been performed. The result shows that the 4D (lightest) gaugino and 
gravitino masses behave as $m_{1/2}$, $m_{3/2} \approx e^{-\pi k R}k$ 
in the large 
warping region $\pi k R \gg 1$ that possesses a large exponential 
suppression, but a negligible effect of the SS twist. 
On the other hand, the twist provides a sizable effect 
in the small warping region $\pi k R < 2$. 

The ratio between the gaugino and the gravitino masses 
$m_{1/2}/m_{3/2}$ 
is suppressed at most by a factor of ${\cal O}(1/10)$ 
within the whole range of warping parameter, $\pi k R$ 
(at least for the simple parameter choice (\ref{eq:simpleex})). 
Then the mediation of SUSY breaking to the gaugino mass 
will be interpreted as a modulus-dominated one (radion 
mediation), and the contribution from 4D chiral compensator 
at the loop level~\cite{Randall:1998uk} is negligible. 
However, in the region $\pi k R \gg 1$, the messenger scale 
can be naturally very low compared to the modulus (radion) 
mediation in a flat space without fine-tuning. 
Then, we may realize a low scale modulus mediation 
without fine-tuning within our framework. 
For more detailed study about the low energy phenomenology, 
the 4D effective theory in $N=1$ superspace would be useful. 
A direct and systematic derivation for such 4D effective 
action starting from the off-shell SUGRA is given 
in Ref.~\cite{Correia:2006pj} based on the $N=1$ 
superfield description of 5D SUGRA proposed in 
Ref.~\cite{PaccettiCorreia:2004ri} and developed 
in Ref.~\cite{Abe:2005ac}. 
An application of these effective action method to the 
present model will be interesting and instructive. 

The orbifold radius remains as a modulus in our simple setup. 
If we introduce, e.g., some boundary induced potential terms 
to stabilize the radion, they can generate constant terms in 
the boundary actions after the radion has a vacuum expectation 
value. The constant superpotentials on the boundaries in our 
model may originate from such constant terms. In such sense, 
the radion stabilization mechanism would have some connection 
to the SUSY breaking structure studied in this paper.

\subsection*{Acknowledgements}
H.~A. and Y.~S. are supported by the Japan Society for the 
Promotion of Science for Young Scientists (No.182496 and No.179241, 
respectively). The numerical calculations were carried out on 
Sushiki at YITP in Kyoto University.

\appendix

\section{Off-shell formulation of 5D orbifold supergravity}
\label{app:offshell}
In this appendix we review the hypermultiplet compensator formulation 
of 5D conformal (off-shell) SUGRA derived in 
Refs.~\cite{Fujita:2001bd,Kugo:2002js}. 

The 5D superconformal algebra consists of 
the Poincar\'e symmetry \mbox{\boldmath $P$}, \mbox{\boldmath $M$}, 
the dilatation symmetry \mbox{\boldmath $D$}, 
the $SU(2)$ symmetry \mbox{\boldmath $U$}, 
the special conformal boosts \mbox{\boldmath $K$}, 
$N=2$ supersymmetry \mbox{\boldmath $Q$}, 
and the conformal supersymmetry \mbox{\boldmath $S$}. 
We use $\mu,\nu,\ldots$ as five-dimensional curved indices 
and $m,n,\ldots$ as the tangent flat indices. 
The gauge fields corresponding to these generators 
$\mbox{\boldmath $X$}_{\!\!A}=
\mbox{\boldmath $P$}_m,\, 
\mbox{\boldmath $M$}_{mn},\, 
\mbox{\boldmath $D$},\, 
\mbox{\boldmath $U$}_{ij},\, 
\mbox{\boldmath $K$}_m,\, 
\mbox{\boldmath $Q$}_i,\, 
\mbox{\boldmath $S$}_i$, 
are respectively 
$h_\mu^{\ A}=
e_\mu^{\ m},\, 
\omega_\mu^{mn},\, 
b_\mu,\, 
V_\mu^{ij},\, 
f_\mu^{\ m},\, 
\psi_\mu^i,\, 
\phi_\mu^i$, 
in the notation of Refs.~\cite{Fujita:2001bd,Kugo:2002js}. 
The index $i=1,2$ is the 
$SU(2)_{\mbox{\scriptsize \boldmath $U$}}$-doublet index 
which is raised and lowered by antisymmetric tensors 
$\epsilon^{ij}=\epsilon_{ij}$. 

In this paper we are interested in the following 
superconformal multiplets: 
\begin{itemize}
\item 
5D Weyl multiplet: 
($e_\mu^{\ m}$, $\psi_\mu^i$, $V_\mu^{ij}$, 
$b_\mu$, $v^{mn}$, $\chi^i$, $D$), 

\item
5D vector multiplet: 
($M$, $W_\mu$, $\Omega^i$, $Y^{ij}$)$^I$, 

\item
5D hypermultiplet: 
(${\cal A}^\alpha_{\ i}$, $\zeta^\alpha$, ${\cal F}^\alpha_{\ i}$). 
\end{itemize}
Here the index $I= 0,1,2,\ldots,n_V$ 
labels the vector multiplets, and $I=0$ component corresponds 
to the central charge vector multiplet\footnote{
Roughly speaking, the vector field in this 
multiplet corresponds to the graviphoton.}. 
For the hypermultiplets, the index $\alpha$ runs as 
$\alpha=1,2,\ldots,2(n_C+n_H)$ where $n_C$ and $n_H$ are the numbers 
of the compensator and the physical hypermultiplets, 
respectively. In this paper we adopt the single compensator case, 
$n_C=1$, and separate the index $\alpha$ such as 
$\alpha=(a,\underline\alpha)$ where $a=1,2$ and 
$\underline\alpha=1,2,\ldots,2n_H$ 
are indices for the compensator and the physical hypermultiplets, 
respectively. 

A superconformal gauge fixing for the reduction to 
5D Poincar\'e SUGRA is given by  
\begin{eqnarray}
\begin{array}{rcl}
\mbox{\boldmath $D$} &:& {\cal N} =M_5^3 \equiv 1, \\
\mbox{\boldmath $U$} &:& {\cal A}^a_{\ i} \propto \delta^a_{\ i}, 
\qquad (n_C=1) \\
\mbox{\boldmath $S$} &:& {\cal N}_I\Omega^{Ii}=0, \\
\mbox{\boldmath $K$} &:& {\cal N}^{-1}\hat{\cal D}_m {\cal N}=0, 
\end{array}
\label{eq:scgf}
\end{eqnarray}
where ${\cal N}=C_{IJK} M^I M^J M^K$ 
is the norm function of 5D SUGRA 
with a totally symmetric constant $C_{IJK}$, 
and ${\cal N}_I=\partial {\cal N}/\partial M^I$. 
The derivative $\hat{\cal D}_m$ denotes the 
superconformal covariant derivative. 
Throughout this paper we take the unit of the 
5D Planck mass, $M_5=1$. 

The off-shell action for 5D SUGRA on $S^1/Z_2$ orbifold 
is given by~\cite{Fujita:2001bd}  
\begin{eqnarray}
S &=& \int d^4x \int dy\, 
\bigg\{ {\cal L}_b + {\cal L}_f + {\cal L}_{\rm aux} 
+\sum_{l=0,\pi} {\cal L}^{(l)} \delta(y-lR) \bigg\}, 
\end{eqnarray}
where ${\cal L}_b$, ${\cal L}_f$, ${\cal L}_{\rm aux}$ 
and ${\cal L}_{N=1}$ are the Lagrangians for the bosonic, 
fermionic, auxiliary fields and the boundary Lagrangian, 
respectively, given by 
\begin{eqnarray}
e^{-1}{\cal L}_b &=& 
-\frac{1}{2}{\cal N}R 
- \frac{1}{4}{\cal N}a_{IJ}F^I_{\mu \nu} F^{\mu \nu J} 
+ \frac{1}{2}{\cal N}a_{IJ} \nabla^m M^I \nabla_m M^J 
+\frac{1}{2} {\cal N}^{IJ} {\cal Y}^{ij}_I {\cal Y}_{Jij}
\nonumber \\ && 
+ \nabla^m {\cal A}^{\bar\alpha}_{\ i} \nabla_m {\cal A}_\alpha^{\ i} 
+ {\cal A}^{\bar\alpha}_{\ i} 
(g^2 M^2)_{\alpha}^{\ \beta} {\cal A}_{\beta}^{\ i} 
+ {\cal N}^{-1}({\cal A}^{\bar\alpha i} \nabla_m {\cal A}_\alpha^{\ j})^2 
\nonumber \\ && 
+ \frac{1}{8} e^{-1} 
c_{IJK} \epsilon^{\lambda \mu \nu \rho \sigma} W^I_\lambda 
\left( F^J_{\mu \nu} F^K_{\rho \sigma}+\frac{1}{2}g[W_\mu, W_\nu]^J
F^K_{\rho \sigma}  +\frac{1}{10}g^2 [W_\mu, W_\nu]^J 
[W_\rho, W_\sigma]^K \right), 
\nonumber \\*[10pt]
e^{-1}{\cal L}_f &=& 
-2i{\cal N} \bar\psi_\mu \gamma^{\mu \nu \rho} \nabla_\nu \psi_\rho 
+2i{\cal N} a_{IJ} \bar\Omega^I \nabla \!\!\!\!/\, \Omega^J 
-2i \bar\zeta^{\bar\alpha} (\nabla \!\!\!\!/\, + gM) \zeta_\alpha 
\nonumber \\ &&
-2i {\cal A}^{\bar\alpha}_{\ i} \nabla_n {\cal A}_{\alpha j} 
\bar\psi_m^{\ (i} \gamma^{mnk} \psi_k^{\ j)} 
+2ia_{IJ} {\cal A}^{\bar\alpha}_{\ i} \nabla_m {\cal A}_{\alpha j} 
\bar\Omega^I_i \gamma^m \Omega^J_j 
-4i \nabla_n {\cal A}^{\bar\alpha}_{\ i} 
\bar\psi_m^i \gamma^n \gamma^m \zeta_\alpha 
\nonumber \\ &&
+{\cal N}(\bar\psi_m \psi_n) 
(\bar\psi_k \gamma^{mnkl} \psi_l + \bar\psi^m \psi^n) 
+ {\cal N}(a_{IJ} \bar\Omega^{Ii} \gamma_m \Omega^{Jj})^2 
-ig{\cal N}_I [\bar\Omega,\Omega]^I 
\nonumber \\ &&
+\frac{1}{8} (\bar\psi_k \gamma^{mnkl} \psi_l 
+ 2 \bar\psi^m \psi^n + a_{JK} \bar\Omega^J \gamma^{mn} \Omega^K 
+ \bar\zeta^{\bar\alpha} \gamma^{mn} \zeta_\alpha)^2 
\nonumber \\ &&
+\frac{i}{4} {\cal N}_I F_{mn}^I (W) 
(\bar\psi_k \gamma^{mnkl} \psi_l + 2 \bar\psi^m \psi^n 
+ a_{JK} \bar\Omega^J \gamma^{mn} \Omega^K 
+ \bar\zeta^{\bar\alpha} \gamma^{mn} \zeta_\alpha) 
\nonumber \\ &&
+i{\cal N}a_{IJ} \bar\psi_m 
(\gamma \cdot F^I(W) -2 \nabla \!\!\!\!/\, M^I) \gamma^m \Omega^J 
\nonumber \\ &&
-2{\cal N}a_{IJ} (\bar\Omega^I \gamma^m \gamma^{nk} \psi_m) 
(\bar\psi_n \gamma_k \Omega^J) 
+2{\cal N}a_{IJ} (\bar\Omega^I \gamma^m \gamma^n \psi_m) 
(\bar\psi_n \Omega^J) 
\nonumber \\ &&
+\frac{1}{4}i{\cal N}_{IJK} \bar\Omega^I \gamma \cdot F^J(W) \Omega^K 
-\frac{2}{3} (\bar\Omega^I \gamma^{mn} \Omega^J \bar\psi_m \gamma_n \Omega^K 
+\bar\psi^i \cdot \gamma \Omega^{Ij} \bar\Omega^J_{(i} \Omega^K_{j)}), 
\nonumber \\*[10pt]
e^{-1} {\cal L}_{\rm aux} &=& 
D'({\cal A}^2+2{\cal N}) 
-8i \bar\chi^{'i} {\cal A}^{\bar\alpha}_{\ i} \zeta_\alpha 
-\frac{1}{2}{\cal N}_{IJ}
(Y^{Iij}-Y^{Iij}_{{\rm sol}})(Y^J_{ij}-Y^J_{ij\,{\rm sol}}) 
\nonumber \\ &&
+2{\cal N}(v-v_{\rm sol})^{mn}(v-v_{\rm sol})_{mn} 
-{\cal N}(V_\mu-V_{{\rm sol}\,\mu})^{ij}(V^\mu-V_{{\rm sol}}^\mu)_{ij} 
\nonumber \\ &&
+\big( 1-W^{I=0}_\mu W^{I=0\,\mu}/(M^{I=0})^2 \big) 
({\cal F}^{\bar\alpha}_{\ i}-{\cal F}^{\bar\alpha}_{\ i\,{\rm sol}}) 
({\cal F}_\alpha^{\ i}-{\cal F}_{\alpha\ {\rm sol}}^{\ i}), 
\label{eq:bulklag}
\end{eqnarray}
and 
\begin{eqnarray}
{\cal L}^{(l)} 
&=& {\textstyle -\frac{3}{2}}
\big[ \phi \bar\phi e^{-K^{(l)}(S,\bar{S})/3} \big]_D
+ \big[ f^{(l)}_{IJ} (S) W^{I \alpha} W^J_\alpha \big]_F
+ \big[ \phi^3 W^{(l)}(S) \big]_F. 
\label{eq:bdrylag}
\end{eqnarray}
Here we have used some expressions defined as 
\begin{eqnarray}
{\cal F}^{\alpha}_{\ i\,{\rm sol}} 
&=& -M^{I=0}(gt_{I=0})^\alpha_{\ \beta}{\cal A}^\beta_{\ i}, 
\nonumber \\
V_{{\rm sol}\,m}^{ij} &=& 
-\frac{1}{2{\cal N}} \left( 
2 {\cal A}^{\bar\alpha (i} \nabla_m {\cal A}_\alpha^{\ j)} 
-i{\cal N}_{IJ} \bar\Omega^{Ii} \gamma_m \Omega^{Jj} \right), 
\nonumber \\
v_{{\rm sol}\,mn} &=& 
-\frac{1}{4{\cal N}} \left\{ 
{\cal N}_I F_{mn}^I(W) -i \left( 
6{\cal N} \bar\psi_m \psi_n 
+ \bar\zeta^{\bar\alpha} \gamma_{mn} \zeta_\alpha 
-\frac{1}{2} {\cal N}_{IJ} \bar\Omega^I \gamma_{mn} \Omega^J 
\right) \right\}, 
\nonumber \\
Y^{Iij}_{{\rm sol}} &=& 
{\cal N}^{IJ} {\cal Y}_J^{ij}, \qquad 
{\cal F}^\alpha_{{\rm sol}\,i} \ = \ 
M^{I=0}(gt_{I=0})^\alpha_{\ \beta} {\cal A}^\beta_{\ i}, 
\nonumber \\
{\cal Y}_I^{ij} &=& 
2 {\cal A}_\alpha^{\ (i} (gt_I)^{\bar\alpha \beta} {\cal A}_\beta^{\ j)} 
+i {\cal N}_{IJK} \bar\Omega^{Ji} \Omega^{Kj}, 
\label{eq:auxsol}
\end{eqnarray}
and $D'$, ${{\chi^i}'}$ are shifted auxiliary fields 
from $D$, $\chi^i$ in the Weyl multiplet respectively 
for which we omit the explicit form (see Ref.~\cite{Fujita:2001bd}). 
The notation $A^{(i}B^{j)}$ is defined as 
$A^{(i}B^{j)}=(A^iB^j+A^jB^i)/2$. 

The bulk gauge kinetic function $a_{IJ}$ is calculated as 
\begin{eqnarray}
a_{IJ} &\equiv& 
-\frac{1}{2} \frac{\partial^2}{\partial M^I \partial M^J} \ln {\cal N}
\ = \ -\frac{1}{2{\cal N}} 
\left( {\cal N}_{IJ} - \frac{{\cal N}_I {\cal N}_J}{{\cal N}} \right). 
\end{eqnarray}
The $n_V+1$ gauge scalar fields $M^I$ are constrained by 
$\mbox{\boldmath $D$}$ gauge fixing shown in Eq.~(\ref{eq:scgf}) 
resulting $n_V$ independent degrees of freedom. 
For hyperscalars ${\cal A}^\alpha_{\ i}$, the notation $\bar\alpha$ 
in the action stands for 
${\cal A}^{\bar\alpha}_{\ i} = 
d_{\beta}^{\ \alpha} {\cal A}^\beta_{\ i}$ 
where 
$d_{\alpha}^{\ \beta} \equiv 
{\rm diag}(\mathbf{1}_{2n_C},-\mathbf{1}_{2n_H})$. 

\begin{table}[t]
\begin{center}
\begin{tabular}{c|c}
\hline 
\multicolumn{2}{l}{\hfill 
\begin{minipage}{0.18\linewidth}
\ \\*[-2pt]
Weyl multiplet \\*[-9pt]
\end{minipage}
\hfill \hspace{0pt}} \\ 
\hline 
$\Pi=+1$ & 
\begin{minipage}{0.65\linewidth}
\ \\*[0pt]
\centerline{
$e_{\underline\mu}^{\ \underline{m}},\,e_y^{\ 4},\,
\psi_{\underline\mu+},\,\psi_{y-},\,\varepsilon_+,\,\eta_-,\,
b_{\underline\mu},\,V_{\underline\mu}^{(3)},\,V_y^{(1,2)},\,
v^{4\underline{m}},\,\chi_+,\,D$} \\*[-5pt]
\end{minipage} 
\\ \hline 
$\Pi=-1$ & 
\begin{minipage}{0.65\linewidth}
\ \\*[0pt]
\centerline{
$e_{\underline\mu}^{\ 4},\,e_y^{\ \underline{m}},\,
\psi_{\underline\mu-},\,\psi_{y+},\,\varepsilon_-,\,\eta_+,\,
b_y,\,V_y^{(3)},\,V_{\underline\mu}^{(1,2)},\,
v^{\underline{m}\underline{n}},\,\chi_-$} \\*[-5pt]
\end{minipage}
\\ \hline \hline 
\multicolumn{2}{l}{\hfill 
\begin{minipage}{0.2\linewidth}
\ \\*[-2pt]
Vector multiplet \\*[-9pt]
\end{minipage}
\hfill \hspace{0pt}} \\ 
\hline 
$\Pi(M)$ & 
\begin{minipage}{0.65\linewidth}
\ \\*[0pt]
\centerline{
$M,\,W_y,\,Y^{(1,2)},\,\Omega_-$} \\*[-6pt]
\end{minipage}
\\ \hline 
$-\Pi(M)$ & 
\begin{minipage}{0.65\linewidth}
\ \\*[0pt]
\centerline{
$W_{\underline\mu},\,Y^{(3)},\,\Omega_+$} \\*[-6pt]
\end{minipage} 
\\ \hline \hline 
\multicolumn{2}{l}{\hfill 
\begin{minipage}{0.18\linewidth}
\ \\*[-2pt]
Hypermultiplet \\*[-9pt]
\end{minipage}
\hfill \hspace{0pt}} 
\\ \hline 
$\Pi({\cal A}^{2\hat\alpha-1}_{\ i=1})$ & 
\begin{minipage}{0.65\linewidth}
\ \\*[0pt]
\centerline{
${\cal A}^{2\hat\alpha-1}_{\ i=1},\,
{\cal A}^{2\hat\alpha}_{\ i=2},\,
{\cal F}^{2\hat\alpha-1}_{\ i=2},\,
{\cal F}^{2\hat\alpha}_{\ i=1},\,
\zeta^{\hat\alpha}_+$} \\*[-6pt]
\end{minipage}
\\ \hline 
$-\Pi({\cal A}^{2\hat\alpha-1}_{\ i=1})$ & 
\begin{minipage}{0.65\linewidth}
\ \\*[0pt]
\centerline{
${\cal A}^{2\hat\alpha-1}_{\ i=2},\,
{\cal A}^{2\hat\alpha}_{\ i=1},\,
{\cal F}^{2\hat\alpha-1}_{\ i=1},\,
{\cal F}^{2\hat\alpha}_{\ i=2},\,
\zeta^{\hat\alpha}_-$} \\*[-6pt]
\end{minipage}
\\ \hline 
\end{tabular}
\end{center}
\caption{The $Z_2$ parity assignment. 
The under-bar means that the index is 4D one. 
The subscript $\pm$ for all the $SU(2)$ Majorana 
spinors is defined as, e.g. 
$\psi_+=\psi^{i=1}_R+\psi^{i=2}_L$ and 
$\psi_-=i(\psi^{i=1}_L+\psi^{i=2}_R)$ where 
$\psi_{R,L}=(1 \pm \gamma_5)\psi/2$, except for 
$\zeta_+^{\hat\alpha}=i(\zeta^{\alpha=2\hat\alpha-1}_L
+\zeta^{\alpha=2\hat\alpha}_R)$ and 
$\zeta_-^{\hat\alpha}=\zeta^{\alpha=2\hat\alpha-1}_R
+\zeta^{\alpha=2\hat\alpha}_L$ 
($\hat\alpha=1,\ldots,n_C+n_H$).}
\label{tab:parity}
\end{table}

It was shown in Ref.~\cite{Fujita:2001bd,Kugo:2002js} that the above 
off-shell SUGRA can be consistently compactified on an orbifold 
$S^1/Z_2$ by the $Z_2$-parity assignment shown in Table~\ref{tab:parity} 
without loss of generality. 
In the boundary action, $\phi$ is the $N=1$ compensator chiral 
multiplet with the Weyl and chiral weight $(w,n)=(1,1)$ induced by 
the 5D compensator hypermultiplet, while $S$ and $W^{I\alpha}$ stand for 
generic chiral matter and gauge (field strength) multiplets 
with $(w,n)=(0,0)$ at the boundaries which consist of either bulk fields 
or pure boundary fields. 
The symbols $[\cdots]_D$ and $[\cdots]_F$ represent the $D$- and 
$F$-term invariant formulae, respectively, in the $N=1$ superconformal 
tensor calculus~\cite{Kugo:2002js,Kugo:1982cu}.
Without the boundary $N=1$ action ${\cal L}_{N=1}$, 
the auxiliary fields on-shell are given by 
$V_\mu=V_{{\rm sol}\,\mu}$, 
$v_{mn}=v_{{\rm sol}\,mn}$, 
$Y^{Iij}=Y^{Iij}_{{\rm sol}}$, 
${\cal F}^\alpha_{\ i}={\cal F}^\alpha_{{\rm sol}\,i}$, 
and ${\cal L}_{\rm aux}$ finally vanishes on-shell, 
$e^{-1}{\cal L}_{\rm aux}^{\rm on \textrm{-} shell}=0$.

\end{document}